\documentclass[doublecol]{epl2-arxiv} 

\usepackage{amsmath}
\usepackage[utf8]{inputenc}

\title{Boundary drive induced formation of aggregate condensates\\ in
  stochastic transport with short-range interactions}
\shorttitle{Boundary drive induced formation of aggregate condensates
  in stochastic transport with short-range interactions}

\author{%
  Hannes Nagel\inst{1}\!\!\!\thanks{E-mail: \email{hannes.nagel@itp.uni-leipzig.de}}
  \and Hildegard Meyer-Ortmanns\inst{2}\!\!\!\thanks{E-mail: \email{h.ortmanns@jacobs-university.de}}
  \and Wolfhard Janke\inst{1}\!\!\!\thanks{E-mail: \email{wolfhard.janke@itp.uni-leipzig.de}}}
\shortauthor{H. Nagel \etal}
\institute{                    
  \inst{1} Institut f\"ur Theoretische Physik, Universit\"at Leipzig,
  Postfach 100~920, 04009~Leipzig, Germany\\
  \inst{2} School of Engineering and Science, Jacobs University
  Bremen, P.O. Box 750561, 28725 Bremen, Germany
}

\pacs{05.60.-k}{Transport processes}
\pacs{02.50.Ey}{Stochastic processes}
\pacs{05.70.Fh}{Phase transitions}

\abstract{
  We discuss the effects of particle exchange through open
  boundaries and the induced drive on the phase structure and
  condensation phenomena of a stochastic transport process with
  tunable short-range interactions featuring pair-factorized steady
  states (PFSS) in the closed system. In this model, the steady state
  of the particle hopping process can be tuned to yield properties
  from the zero-range process (ZRP) condensation model to those of
  models with spatially extended condensates. By varying the particle
  exchange rates as well as the presence of a global drift, we observe
  a phase transition from a free particle gas to a phase with
  condensates aggregated to the boundaries. While this transition is
  similar to previous results for the ZRP, we find that the mechanism
  is different as the presence of the boundary actually influences the
  interaction due to the non-zero interaction range.}

\begin{document}

\maketitle


Condensation phenomena are observed in a broad range of physical
processes. While they are originally associated with phase transitions
of matter from the gas state to some liquid or solid state, they are
also closely related to nucleation and coarsening phenomena. Examples
of condensation appear in processes such as the formation of breath
figures~\cite{Beysens1986}, Bose--Einstein condensation
\cite{Bialas1996}, polymer aggregation~\cite{Zierenberg2014}, but in a
wider sense also in more generic systems like networks as the
formation of clusters \cite{Dorogovtsev:networks} through the
accumulation of links on sites.

For many such systems, the involved condensation process can be
modeled as a stochastic transport process with a set of particles
occupying a number of discrete sites. With particles representing
microscopic to macroscopic objects and appropriate dynamics a wide
spectrum of physical processes has been studied. Examples include
refs.~\cite{Bialas1996, Dorogovtsev:networks} mentioned above, but also
processes such as wealth condensation \cite{Burda2002} or traffic flow
\cite{Kaupuzs2005}.


The zero-range process (ZRP) with condensation
dynamics~\cite{Evans2000,Evans2005} is a well known paradigm of such
transport processes. While it has a fully symmetric steady state,
above some critical density $\rho_{\text{c}}$ the symmetry breaks
spontaneously and a particle condensate emerges at a single site such
that the density at the remaining sites stays critical. When
short-range interactions are introduced, a similar condensation
process can be observed with the main difference, that condensates can
be spatially
extended~\cite{Evans2006,Waclaw2009c,Waclaw2009b,Ehrenpreis2014}. In
this work, however, we shall consider condensation not as an effect in
the steady state, but as a signature of a boundary induced phase
transition. Such transitions can occur in driven systems, where the
drive is implemented in terms of the interaction at the boundary of
the system~\cite{Popkov1999}. A well known transport process with such
a transition is the totally asymmetric simple exclusion process
(TASEP)~\cite{Spitzer1970,Blythe2004}, where a high-density, a
low-density and a maximal-current phase exist~\cite{Krug1991}. For the
ZRP, such effects of open boundaries and driven particle exchange have
been studied using analytical and numerical methods by Levine
\etal~\cite{Levine2005}.

While the interaction of particles with the boundary of the ZRP is
merely the injection and removal of particles, the presence of the
boundary influences hopping at nearby sites once short-range
interactions are introduced. Within this paper, we use the term
\emph{open} boundaries in the sense that the boundary sites only have
one interaction bond towards the bulk of the system. While this is a
natural choice for an isolated system and consistent with
ref.~\cite{Levine2005}, it is also conceivable that the boundary sites
interact with a mean-field occupation outside the system, for example
to model a compartment of a large system separated by membranes.

\begin{figure}
  \centering
  \onefigure[width=0.9\columnwidth]{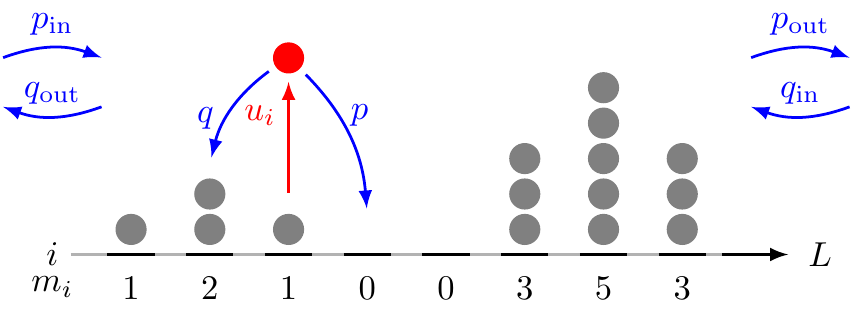}
  \caption{Schematic representation of the dynamics of a particle
    hopping process on a one-dimensional lattice with $L$ sites,
    hopping rate $u_{i}$ and drift parameters $p$, $q$. At the
    boundary sites $i=L$ and $1$ they are replaced by the removal
    parameters $p_{\text{out}}$ and $q_{\text{out}}$, respectively.
    Particle injection (rate parameters
    $p_{\textrm{in}},q_{\textrm{in}}$) is independent of the hopping
    rate.}
  \label{fig:scheme}
\end{figure}

We consider a one-dimensional lattice with $L$ sites and a gas of
indistinguishable particles. Each site $i$ can hold an arbitrary
number $m_{i}$ of particles. Particles can hop between sites as well
as enter and exit the system through the boundaries. The dynamics in
terms of a discrete stochastic process consists of two steps as
indicated in fig.~\ref{fig:scheme}: First, a particle may leave from a
randomly chosen site $i$ with probability proportional to the hopping
rate $u_{i}$. Second, the particle hops to the left with rate $q$ or
to the right neighbor with rate $p$. This allows the implementation of
symmetric
($p=q=1/2, p_{\text{in}}=q_{\text{in}},
p_{\text{out}}=q_{\text{out}}$)
as well as partially and totally asymmetric
($p=1, q=q_{\text{in}}=q_{\text{out}}=0$) hopping dynamics. At the
boundary sites $i=L$ and $1$, the drift parameters $p$ and $q$ are
replaced by the removal parameters $p_{\text{out}}$ and
$q_{\text{out}}$ which enter into the rates of particle removal
$u_{L}p_{\text{out}}$ and $u_{1}q_{\text{out}}$, respectively. The
injection rates are independent of the occupation numbers and hence
directly given by the parameters $p_{\text{in}}$ and $q_{\text{in}}$.
Due to this explicit particle injection and removal through the open
boundaries, the total number of particles $M(t)=\sum_{i=1}^{L}m_{i}$
is not conserved. For the hopping rate function, we consider the form
\begin{equation}
  u_{i} = \prod_{\langle i,j \rangle} \frac{g(m_{i}-1, m_{j})}{g(m_{i},m_{j})},
  \label{eq:hopping-rates}
\end{equation}
with a symmetric, non-negative weight function $g(m,n)$ given for each
bond $\langle i,j \rangle$ of the lattice.\footnote{This form of the
  hopping rate with fixed $i$ implicitly realizes the fact, that there
  is no interaction beyond the boundary sites.} Whereas in a closed
system with the total number of particles conserved this choice leads
to a steady state~\cite{Evans2006,Waclaw2009b} of the form
\begin{equation}
  \label{eq:steady-state}
  P_{M,L}(\{m\}) = 
  \frac{1}{Z_{M,L}} \prod_{\langle i,j\rangle} g(m_{i},m_{j})\delta_{\sum_{i=1}^{L}m_{i}, M},
\end{equation}
this is not necessarily the case with open boundaries. However, since
the steady state of the closed system is rooted in the factorization
property over bonds, i.e., pairs of sites, one usually still refers to
it as pair-factorized steady state (PFSS). The normalization constant
$Z_{M,L}$ in the steady state \eqref{eq:steady-state} has the same
function as the partition function in an equilibrium model.

\begin{figure}
\centering
  \includegraphics{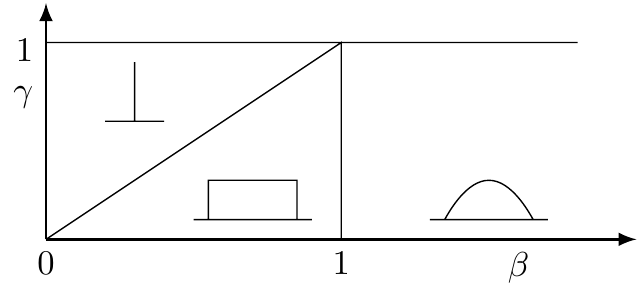}
  \caption{Regimes of different condensate shapes of the considered
    tunable model: No condensation for $\gamma>1$, single-site
    condensates for $\beta < \gamma \le 1$, rectangular condensates
    for $\gamma < \beta < 1$ and smooth parabolic condensates for
    $\beta > 1$ and $\gamma \le 1$.}
  \label{fig:sketch-tunable}
\end{figure}

The weight function $g(m,n)$ gives the interaction between
particles. Here, we consider the tunable weights
\begin{equation}
  g(m,n) = 
  \exp\left[ -\left\vert m-n \right\vert^{\beta} - \frac{1}{2} (m^{\gamma} + n^{\gamma}) \right]
  \label{eq:weights}
\end{equation}
proposed by Wac{\l}aw \etal~\cite{Waclaw2009c,Waclaw2009b}. By tuning
the parameters $\beta,\gamma$, properties of the condensation process
such as the critical density and the condensate's shape and extension
can be chosen~\cite{Ehrenpreis2014}. This allows us to study the model
in a regime with strong nearest-neighbor interactions, where spatially
extended condensates occur in the steady state model ($\gamma \le 1$
and $\gamma<\beta<1$ for rectangular and $\beta>1$ for smooth
condensates), while still being able to directly compare to the
results of Levine \etal~\cite{Levine2005} by tuning the model towards
the weak nearest-neighbour interaction regime
($\beta<\gamma \le1$), where its properties are similar to the ZRP
considered there. Figure~\ref{fig:sketch-tunable} summarizes the
phases of this model with periodic boundaries.

\begin{figure}[t]
  \onefigure{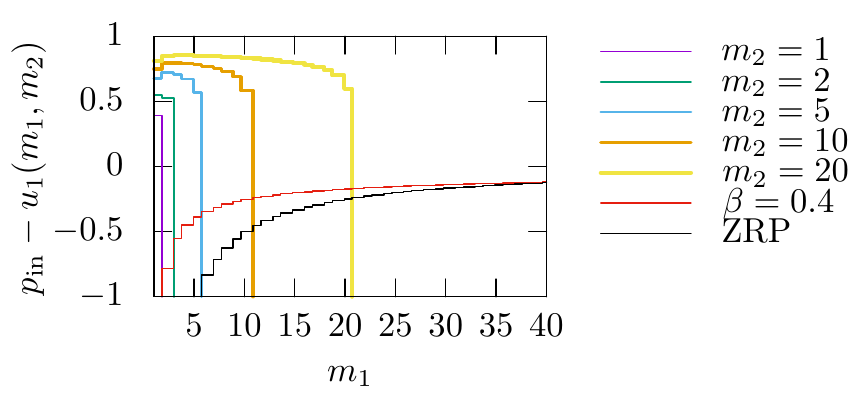}
  \caption{Drift $p_{\text{in}} - u_{1}(m_{1},m_{2})$ of the
    occupation number of the first site, $m_{1}$, for different
    occupations at the second site, $m_{2}$, for our model
    ($\beta=1.2, \gamma=0.6$) at $p_{\text{in}}=1$ as well as for weak
    short-range interactions ($\beta=0.4,\gamma=0.6, m_{2}=0$) and the
    ZRP with totally asymmetric ($p=1, q=0$) dynamics. For
    $\beta=0.4$, there is no significant dependence on the second site
    occupation $m_{2}$, for the ZRP there is none.}
  \label{fig:drift}
\end{figure}

From the research in ref.~\cite{Levine2005} we know that the ZRP
condensation model with hopping rates $u(m)=1+b/m$ in the condensation
regime $b>2$ has a phase transition induced by boundary drive. A
homogeneous gas phase where a steady state still exists and one
(symmetric hopping) or two (totally asymmetric hopping) phases with
single-site condensates aggregated to the boundary are observed. This
transition occurs for $p_{\text{in}}>p_{\text{out}}$ for symmetric and
$p_{\text{in}}>1$ for totally asymmetric hopping. For the ZRP, the
presence of the boundary does not affect the interaction at the
boundary sites other than by injection or removal of particles. This
makes it possible to understand the condensate formation at a boundary
site for totally asymmetric hopping by only considering the difference
of the fluxes into and out of that site~\cite{Levine2005}. The
occupation number of the boundary site is then treated as a biased
bounded random walk with drift equal to this net flux. As shown in
fig.~\ref{fig:drift}, this drift becomes positive for sufficiently
high influx rate $p_{\text{in}}>1$ and occupation $m_{1}$ so that a
condensate can emerge and grow after a sufficiently large fluctuation
of $m_{1}$.
For the ZRP, the effect of these boundary condensates to the
bulk system is then, that they act as particle reservoirs that hold
the particle density in the bulk stationary at the critical density
$\rho_{\text{c}}=1/(b-2)$.
%
%
It is also apparent from fig.~\ref{fig:drift} that with short-range
interactions this drift behaves entirely differently and also depends
on the second site's occupation $m_{2}$. We will pick up this
observation further below, when we discuss the corresponding phase of
aggregate condensate formation for the short-range interaction model.


In our own preliminary work~\cite{Nagel2014} we additionally observed
for the PFSS model the formation of condensates in the bulk system in
the aggregate condensate phase for symmetric hopping. From this we
could also assume that with short-range interactions the phase
diagram induced by particle exchange and external drive is
similar to that of the ZRP. However, there were limitations of the
simulation method of the stochastic process because no upper bounds
exist for the hopping rate function generated by the
weights~\eqref{eq:weights}, so that we were not yet able to study the
phase diagram with short-range interactions to our
satisfaction.

To avoid these problems, we used here an improved rejection-free
kinetic Monte Carlo (KMC) method, 
originally introduced for the simulation of coupled rate equations in
chemical systems by Gillespie~\cite{Gillespie1976,Gillespie1977}. This
allows us to directly use the interaction rates of the system which
makes simulation much more efficient in situations where no steady
state exists.

\begin{figure}
  \onefigure[width=0.95\columnwidth]{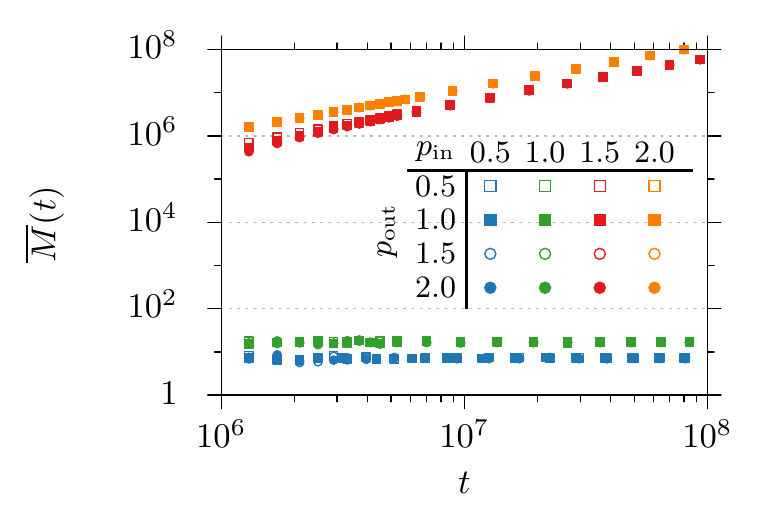}
  \caption{The total number of particles $M(t)$ in the system versus
    time. For sufficiently large times, a stationary base line is a
    signature of the gas phase while linear growth occurs in the
    aggregate condensate phase. Curves averaged from 25 independent
    simulations at
    $\beta=1.2, \gamma=0.6, p=q=1/2, L=256$.}
  \label{fig:totalmass}
\end{figure}
\begin{figure}
  \onefigure{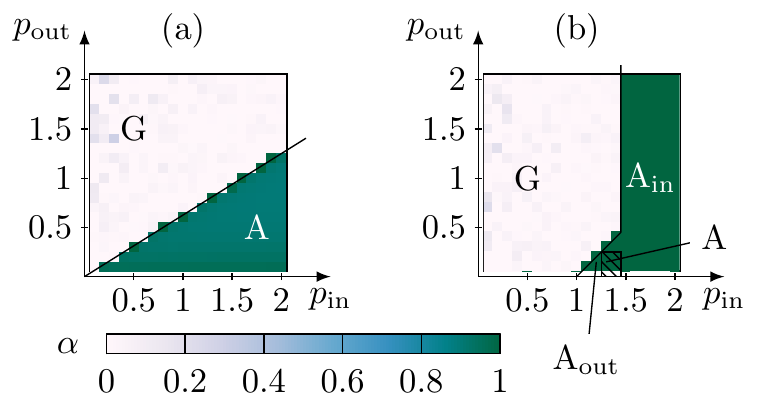}
  \caption{Scaling exponent $\alpha(p_{\text{in}},p_{\text{out}})$ of
    the total mass $M(t)$ over time for (a) symmetric
    ($p=q=1/2, p_{\text{in}}=q_{\text{in}},
    p_{\text{out}}=q_{\text{out}}$)
    and (b) totally asymmetric
    ($p=1, q=q_{\text{in}}=q_{\text{out}}=0$) dynamics estimated at
    large times $10^{7}<t<10^{8}$ for $\beta=1.2,\gamma=0.6$.}
  \label{fig:phases-alpha}
\end{figure}

In the following we will outline how we identify the phase
diagram. As a main observable to indicate the expected
phase transition we measure the total number of particles
$M(t)$ as shown for symmetric hopping in
fig.~\ref{fig:totalmass} and estimate the scaling exponent
$\alpha$ under the assumption that
$M(t) \propto t^{\alpha}$ grows as some power of time due
to absorption of a fraction of particles entering the system. We can
thus identify regions in $(p_{\text{in}},p_{\text{out}})$, where the
system continually absorbs particles and the steady state breaks down
in the regions where $\alpha \approx 1$. This is demonstrated in
fig.~\ref{fig:phases-alpha}, which shows the average total
number of particles versus time for several combinations of particle
exchange rates $p_{\text{in}}$ and $p_{\text{out}}$.

\begin{figure}
  \onefigure{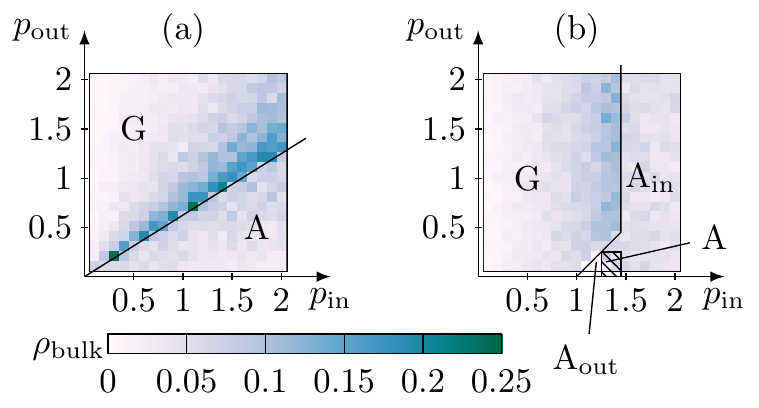}
  \caption{Bulk density
    $\rho_{\text{bulk}}(p_{\text{in}}, p_{\text{out}})$ for (a)
    symmetric
    ($p=q=1/2, p_{\text{in}}=q_{\text{in}},
    p_{\text{out}}=q_{\text{out}}$)
    and (b) totally asymmetric
    $p=1, q=q_{\text{in}}=q_{\text{out}}=0$) dynamics with interaction
    parameters $\beta=1.2, \gamma=0.6$. The critical density observed
    in a closed system of comparable size $L=M=256$ for
      these parameters is $\rho_{\text{crit}}\approx 0.3$.}
  \label{fig:phases-bulkdensity}
\end{figure}

\begin{figure*}
  \centering
  (a)\includegraphics{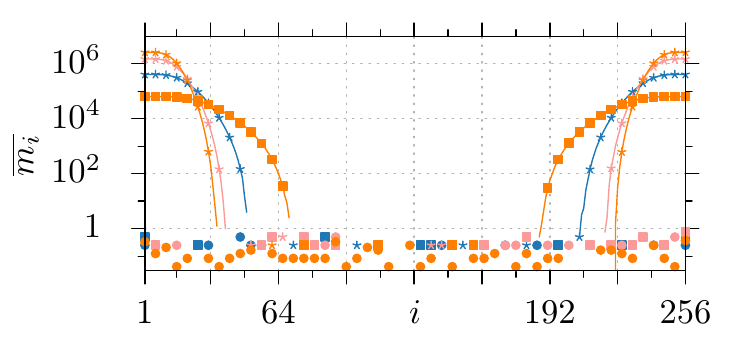}
  \hfill
  (b)\includegraphics{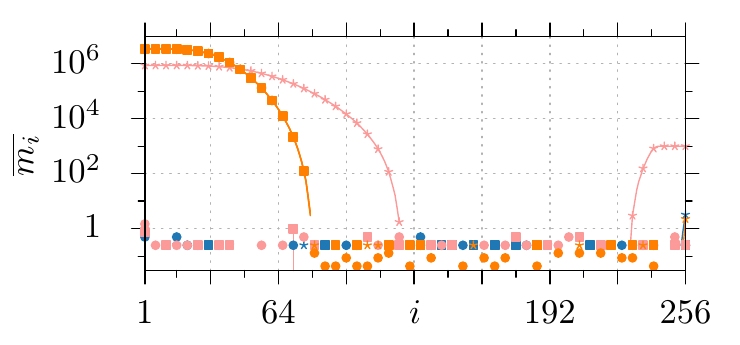}
  \includegraphics{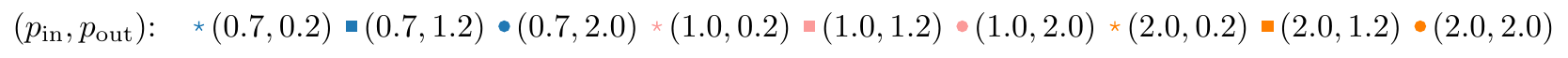}
  \caption{Average occupation number profile for (a) symmetric hopping
    ($p=q=1/2, p_{\text{in}}=q_{\text{in}},
    p_{\text{out}}=q_{\text{out}}$)
    and (b) totally asymmetric hopping
    ($p=1,q=q_{\text{in}}=q_{\text{out}}=0$) for varying particle
    influx and outflux rates. The interaction parameters are
    $\beta=1.2, \gamma=0.6$ and the system is of size $L=256$.}
  \label{fig:profiles}
\end{figure*}

In contrast, for sufficiently small values of the influx rate
$p_{\text{in}}$ we obtain $\alpha\approx 0$ which reflects that the
number of particles in the system remains stationary and suggests that
a steady state exists. In the latter case, the particles are
homogeneously distributed in the system as a thin gas, rarely
interacting (phase~G). This is visible in the low average occupation
number density $\rho_{\text{bulk}}$ in
fig.~\ref{fig:phases-bulkdensity} as well as in the average
occupation profile shown in fig.~\ref{fig:profiles} (flat
profiles). In the gas phase the overall density $\rho$ is identical to
the bulk density where any particle accumulations at the boundaries
are neglected. In the regions with positive $\alpha$ (dark regions in
fig.~\ref{fig:phases-alpha}), particles accumulate in the
system. This occurs through the emergence of spatially extended
condensates at the boundaries, as can be seen in
fig.~\ref{fig:profiles}. As in this regime the condensates
are bound to the boundary, we refer to it as the aggregate condensate
phase (A). While for symmetric hopping
($p=q=1/2$) identical formations appear at both
boundaries [fig.~\ref{fig:profiles}(a)], distinctly shaped
and independent condensates are observed for totally asymmetric
hopping ($p=1, q=0$)
[fig.~\ref{fig:profiles}(b)]. These condensates in the
latter case appear separately in the regions $\text{A}_{\text{in}}$
and $\text{A}_{\text{out}}$ or together in the region A of
figs.~\ref{fig:phases-alpha} and~\ref{fig:phases-bulkdensity}.

At the boundary with particle influx, for a sufficiently high
injection rate $p_{\text{in}}$ in the regime $\text{A}_{\text{in}}$,
an aggregate condensate forms that obeys the same envelope shape as a
bulk condensate in the periodic model~\cite{Waclaw2009b} [both
boundaries in fig.~\ref{fig:profiles}(a), left boundary in (b)]. There
also appears a regime $\text{A}_{\text{out}}$, where particles
condense prior to leaving the system, as well as a coexistence region
$\text{A}$ [for example $p_{\text{in}}=1,p_{\text{out}}=0.2$ in
fig.~\ref{fig:phases-bulkdensity}(b)]. For symmetric hopping the
inbound condensate coexists with the outbound condensate, the former,
however, dominates the shape due to its much larger contribution. The
transition lines from the gas phase G to the aggregate condensate
phase A for the coupling constants $\beta=1.2, \gamma=0.6$ are found
at $p_{\text{in}} > 0.625 p_{\text{out}}$ for symmetric and
$p_{\text{in}} > p_{\text{in,crit}} = 1.43 \pm 0.02$ for totally
asymmetric hopping as shown in figs.~\ref{fig:phases-alpha}
and~\ref{fig:phases-bulkdensity}. We also observe that the location of
the transition lines depends strongly on the couplings $\beta, \gamma$
as well as the type of interaction at the boundaries. For instance, by
decreasing the strength of the short-range interactions, the critical
lines approach those observed for the ZRP until they match for
$\beta < \gamma$. An interesting modification of the interaction at
the boundaries that consists of adding an explicit bond from the
boundary sites to an outer site with occupation $m_{\infty}$,
effectively introducing fixed boundary conditions, leads to a shift of
the transition lines towards larger values of $p_{\text{out}}$ for
asymmetric hopping, so that the phases $\text{A}_{\text{out}}$ and A
become significant regions of the phase
diagram~\cite{Nagel2015InPrep}.

While the phase diagrams given in figs.~\ref{fig:phases-alpha}
and~\ref{fig:phases-bulkdensity} are similar to those of the
ZRP~\cite{Levine2005,Nagel2014} with comparable boundary conditions,
we do observe qualitative differences. With short-range interactions,
the bulk density of the gas phase increases towards the critical
density while approaching the transition line towards aggregate
condensate formation (see fig.~\ref{fig:phases-bulkdensity}). However,
while the occupation density in the bulk system, apart from the
aggregate condensates, remains at criticality for the ZRP as well as
weak short-range interactions $\beta < \gamma$, we observe a sharp
decline at the transition line to a bulk density considerably smaller
than the critical density here.

On a higher level, with significant short-range interactions
\eqref{eq:weights}, i.e.\ $\beta>1$, the formation of the aggregate
condensate at the boundary sites for asymmetric dynamics cannot simply
be understood by means of a bounded, biased random walk of the first
site's occupation number as with the ZRP. Only with weak
nearest-neighbour couplings $\beta<\gamma$ in our considered model
this explanation remains valid, as suggested by the monotony and limit
of the drift shown in fig.~\ref{fig:drift}. With
short-range interactions such a drift also depends on the occupation
of the neighbour site. As shown in fig.~\ref{fig:drift},
for $\beta>1$ this drift is positive even below the observed critical
particle influx rate $p_{\text{in}}<p_{\text{in,crit}}\approx 1.43$
for $1 \le m_{1} \le m_{2}$ but becomes strongly negative for
$m_{1}>m_{2}$. That is, no fluctuation of the first sites occupation
alone can lead to growth of the aggregate condensate in this
approach. However, the region in which the drift remains positive
expands for higher occupations of the second site: The aggregate
condensate can only emerge and exist as a spatially extended
condensate because a sufficiently high occupation at a couple of the
first sites of the system is required to result in positive drift of
the first site's occupation number $m_{1}$.

A second mechanism of the emergence of the aggregate condensate is
only observed with short-range interactions. For the ZRP or
$\beta<\gamma$ the condensate directly forms at the boundary
site. With strong short-range interactions $\beta>1$, however, the
condensate may initially emerge in the bulk system and then relatively
fast connect to the boundary.



We observe boundary drive induced phase separation that becomes
visible as the formation of aggregated condensates at the system
boundaries from a homogeneous gas phase. The phase diagram itself is
similar to that of the ZRP as discussed in ref.~\cite{Levine2005},
specifically the properties of the homogeneous gas phase where the
steady state is not broken are much alike. The observed aggregate
condensates are spatially extended as expected. The accumulation of
particles in these condensates is much stronger than in the ZRP or
than expected, decreasing the particle density in the bulk system far
below the critical density of the steady state model, therefore
excluding the formation of stable droplets in the bulk system.

Since with short-range interactions nearby boundaries do affect the
dynamics, it is worthwhile to also look at the influence of the
specific implementation of the boundaries. One approach to this could
be to assume virtual bonds across the boundaries to a mean-field
occupation $m_{\infty}$ as suggested above, so that the homogeneity of
the system is not a priori broken by the missing bonds. Although this
is a large change of the interaction strengths at the boundary, we
merely observe different positions of the phase boundaries as well as
a deformation of the shape of the aggregate condensate as it couples
to the mean-field occupation $m_{\infty}$ beyond the boundary. A more
interesting change is to make particle injection and removal
symmetric. With the rate of particle removal set constant to
$p_{\text{out}}$, we observe an additional phase with a steady state
for symmetric hopping in between the gas and aggregate condensate
phases. Most notably, it features a single large condensate that
extends the complete bulk of the system but falls to zero nearby the
boundaries. For a more detailed discussion of these observations, we
refer to ref.~\cite{Nagel2015InPrep}.


To conclude, we numerically studied the emergence of phase transitions
induced by driven particle exchange through open boundaries in a
stochastic transport process with tunable short-range interaction. We
observed a gas phase and two aggregate condensate phases, where
particles accumulate at the boundaries. We presented phase diagrams
for the extended smooth condensate regime of the model and discussed
the phases' properties.  We compared these observations with
analytical and numerical results for the zero-range process by Levine
\etal~\cite{Levine2005} where similar phases are
observed. While the gas phases in both systems are
  qualitatively identical, there are noteworthy differences in the
  aggregate condensate phases such as the spatial extension and in the
  case of asymmetric hopping different shapes of the aggregate
  condensates and the below critical bulk density. Furthermore, we
  found that the mechanism of the formation of aggregate condensates
  is different as well as supplemented by an additional mechanism,
  where the condensate forms in the bulk and quickly aggregates at the
  boundary. Finally, we shortly discussed other types of open
  boundaries that we deem important for a systematic study of a
  short-range interaction model with condensation where we observed
  additional phases featuring a large bulk condensate and a
  homogeneous fluid, respectively. Therefore, additionally to the
  boundary drive that our model shares with the ZRP
  model~\cite{Levine2005}, the specific interaction with the boundary
  strongly influences the observable phenomena to the point where
  extra phases become observable.

%


As an outlook to future work we see many possibilities for further
research. For example, when we considered \emph{constant removal} at
the boundary to achieve a symmetry between particle injection and
removal, one might additionally take the other choice and use
\emph{hopping injection}, i.e., particles must successfully hop into
the system. Another intriguing variation is to actually implement the
systems considered here embedded in a larger, possibly periodic,
system and interpret the boundaries of the inner system as membranes.

\acknowledgments We would like to thank the DFG (German Science
Foundation) for financial support under the twin Grants No.\
JA~483/27-1 and ME~1332/17-1.  We further acknowledge
support by the DFH-UFA graduate school under Grant No.\ CDFA-02-07.



\end{document}